\documentclass[journal=nalefd]{achemso}

\usepackage{units}
\usepackage{amsmath}
\usepackage{amssymb}
\usepackage[super]{nth}

\setkeys{acs}{maxauthors=10, etalmode=truncate}

\title{Fano Resonance in an Electrically Driven Plasmonic Device}
\author{Yuval Vardi}
\email{yuval.vardi@weizmann.ac.il}
\altaffiliation{Contributed equally to this work}

\author{Eyal Cohen-Hoshen}
\altaffiliation{Contributed equally to this work}

\author{Guy Shalem}
\author{Israel Bar-Joseph}

\affiliation{Department of Condensed Matter Physics, Weizmann Institute of Science, Rehovot 76100, Israel}

\begin{document}

\renewcommand*{\natmovechars}{}

\begin{abstract}
We present an electrically driven plasmonic device consisting of a gold nanoparticle trapped in a gap between
two electrodes. The tunneling current in the device generates plasmons,
which decay radiatively. The emitted spectrum extends up to an energy that depends on the applied voltage.
Characterization of the electrical conductance at low temperatures allows us to extract the voltage drop on each tunnel barrier
and the corresponding emitted spectrum. In several devices we find a pronounced sharp asymmetrical dip in the spectrum,
which we identify as a Fano resonance.
Finite-difference time-domain (FDTD) calculations reveal that this resonance is due to interference between the nanoparticle and electrodes
dipolar fields, and can be conveniently controlled by the structural parameters.


\end{abstract}

Electrically driven plasmonic devices may offer unique opportunities
as a research tool and for practical applications\cite{GarciadeAbajo2010,Fan2012,Li2013,Rai2013,Sheldon2014}.
In such devices, current that flows across a metallic tunnel junction
excites a plasmon, which gives rise to light emission. The generation
of the plasmon at the tunnel junction\cite{Savage2012} is equivalent
to feeding an antenna at its source. In that sense, it is different
from regular optical excitation of plasmons, where the far field illumination
is coupled to the plasmonic antenna, and then coupled out. This local
nature of the excitation allows easy access into evanescent (or "dark")
modes, which are not easily excited by far field illumination\cite{Hao2008,Savage2012}.
From a more practical point of view, such devices may pave the way
for easy realization of on-chip optical communication and sensors\cite{Bendana2011,Jeong2013,Huang2014,Sheldon2014,Cox2014,Tielrooij2015}.

The possibility to generate light in tunnel junctions was first suggested
and demonstrated more than four decades ago. Light emission in metal-oxide-metal
planar tunnel junctions was observed and was attributed to plasmon
mediated tunneling\cite{Lambe1976,Hansma1978,Hone1978,Rendell1981}.
A renewed interest in this phenomenon occurred in the early nineties,
when light emission was reported in STM experiments\cite{Coombs1988,Gimzewski1989,Berndt1991}.
The excellent control over the tunneling barrier in this system allowed
detailed investigation of the light emission process. Recently, this
concept was implemented in planar plasmonic structure, consisting
of a metallic nanoparticle between two larger electrodes\cite{Kern2015}.

\begin{figure}
\includegraphics[width=1\columnwidth]{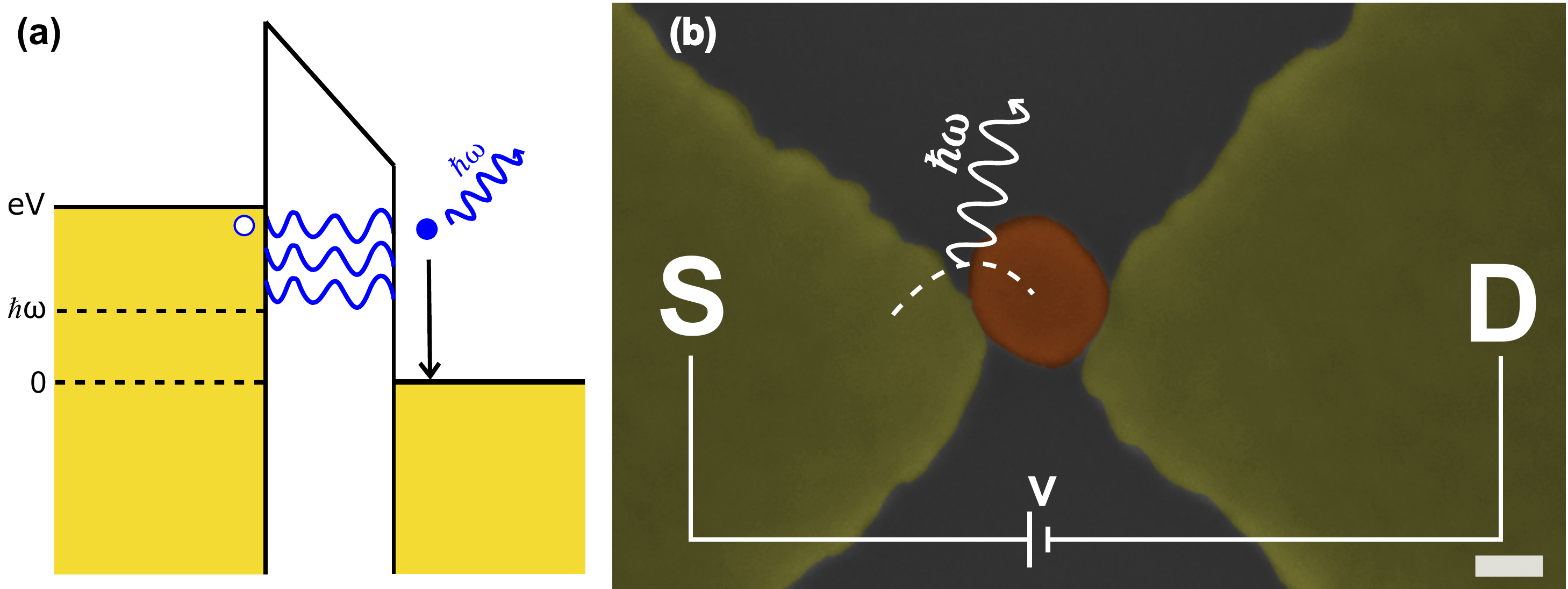}
\caption{(a) Energy scheme for tunneling across a single metallic tunnel junction
with a bias voltage $V$. Light emission at energy $\hbar\omega$
is generated from tunneling electrons with energy $\hbar\omega<E<eV$.
(b) Scanning electron microscope image (false color) of the SET -
a trapped gold nanoparticle between source (S) and drain (D) electrodes.
Tunneling current produces light emission. The scale bar corresponds
to $\unit[25]{nm}$.}
\label{fig1}

\end{figure}

The mechanism of light emission in such structures is well understood.
Consider a metallic tunnel junction, biased by a voltage $V$, as
depicted in Figure~\ref{fig1}(a). During the tunneling process electrons
may lose a fraction or all of their initial energy $eV$, exciting
a plasmon, and consequently emitting photons. One may express the
emitted power as
\begin{equation}
S\left(\omega\right)=\left|I_{\omega}\right|^{2}*g\left(\omega\right)\label{eq:general_emission_spectrum}
\end{equation}
 where $g\left(\omega\right)$ is the plasmonic spectrum, and $\left|I_{\omega}\right|^{2}$
is the power spectrum of the generator current, which can be expressed
as $\left|I_{\omega}\right|^{2}=\sum_{i,f}\left|\langle f|\hat{T}|i\rangle\right|^{2}\delta\left(\hbar\omega-\left(E_{f}-E_{i}\right)\right)$,
where $\hat{T}$ represents the tunneling matrix element operator.
It is evident that at the limit of energy independent tunneling and
zero temperature one can approximate this expression by\cite{Hone1978,Rendell1981,Persson1992,Uehara1992}

\begin{equation}
S\left(\omega\right)=\begin{cases}
\left(\frac{1}{2\pi R_{0}}\right)\left(eV-\hbar\omega\right)g\left(\omega\right) & 0<\hbar\omega<eV\\
0 & eV<\hbar\omega
\end{cases}\label{eq:emission_spectrum}
\end{equation}
where $R_{0}$ is the DC junction resistance, and the applied voltage
$V$ determines the maximal emitted photon energy. In practice, the
tunneling rate does depend on energy, and the linear dependence on
voltage is a good approximation only for $\hbar\omega\sim eV$.
Furthermore, it was shown that photon emission at energies larger than $eV$ can also be obtained.
This is attributed to high order processes, in which an electron relaxation is accompanied by an excitation
of another electron to an energy above the Fermi level\cite{Schneider2013,Kaasbjerg2015,Buret2015}.

In this paper we use a single electron transistor (SET) as a light
emitting device. The general structure of the device is shown in Figure~\ref{fig1}(b):
electrons tunnel from the source electrode (S) to the nanoparticle
(NP) and then into the drain electrode (D). We show that the use of
this structure allows us to properly characterize the electrical properties
of the two tunnel barriers, and determine their role in the light
emission process. In some devices we find a Fano resonance, resulting
from interference between the nanoparticle and electrodes dipolar fields.
This resonance is seen due to the local nature of the excitation,
and is manifested as a sharp asymmetrical spectral dip. We show that
the spectral position of this resonance can be conveniently controlled
by the design of the structural parameters.

We have previously demonstrated a technique of preparing metallic
NP-based SET devices\cite{Guttman2011,Vardi2014}. A bow-tie electrode
structure is patterned by e-beam lithography on n-doped silicon substrate
covered by a $\unit[100]{nm}$ SiO$_{2}$ layer. Gold NPs of $\unit[55]{nm}$
diameter are chemically synthesized\cite{Cohen-Hoshen2012} and covered
by a dense capping layer of mercaptosuccinic acid. The thickness of
this layer is $\sim\unit[0.5]{nm}$, defining a sub-nanometer tunnel
barrier between the NP and the electrodes. The NPs are brought into
a gap of $\unit[30-35]{nm}$ between the two electrodes by means of
electrostatic trapping\bibnotemark[2]. This provides an easy process,
by which several devices can be simultaneously prepared. Optical and
electrical measurements were conducted in vacuum inside an optical
cryostat (Janis Research ST-500HT), both at room temperature and at
cryogenic temperatures ($T=\unit[4.2]{K}-\unit[6]{K}$).

\begin{figure}
\includegraphics[width=0.9\columnwidth]{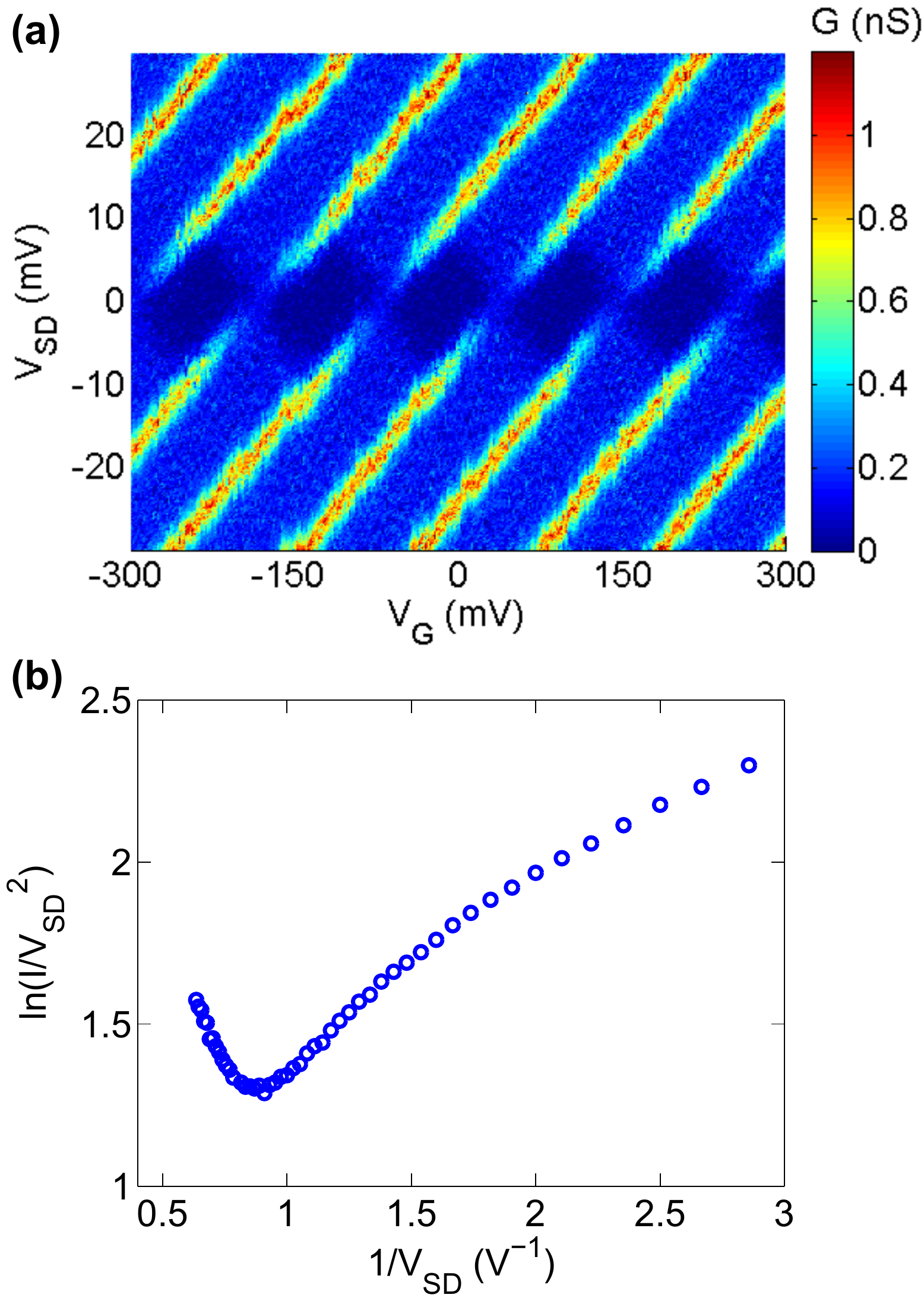}
\caption{(a) Measurement of the differential conductance ($G=dI/dV_{SD}$)
of a device as a function of the back-gate voltage $V_{G}$ and the
source-drain bias voltage $V_{SD}$, at temperature of $\unit[4.2]{K}$.
(b) Measured current-voltage ($I-V_{SD}$) relation presented at the
Fowler-Nordheim representation, where the natural logarithm of $I/V_{SD}^{2}$
is plotted as a function of the inverse voltage $V_{SD}^{-1}$.}
\label{fig2}
\end{figure}

Figure \ref{fig2}(a) shows the measured differential conductance
of a typical device at the $V_{SD}-V_{G}$
plane, where $V_{SD}$ and $V_{G}$ are the source-drain and gate
voltages, respectively. The characteristic diamonds structure consisting
of low conductance Coulomb blockade regions is clearly seen near $V_{SD}=0$.
As we move along the $V_{G}$ axis, each diamond corresponds to charging
the NP by an additional electron. It is evident that this structure
is highly asymmetric: strong conductance peaks appear at the upper
left and lower right boundaries of the diamonds, and they are much
weaker at the opposite sides. This asymmetry becomes more visible
at higher voltages, where it is seen as parallel conductance lines.
The origin of this asymmetry is the difference in tunnel conductance
between the two barriers. By fitting the measurement to a conductance
model\cite{Guttman2011,Vardi2014} we could extract the relevant parameters
characterizing each barrier\bibnotemark[2], and in particular the
tunnel conductance $g_{1}$ and $g_{2}$ of the two barriers. We note
that the ratio between these two conductances determines the division
of $V_{SD}$ between the two barriers. In this particular device we
obtain that $g_{1}=\unit[0.3]{nS}$ and $g_{2}=\unit[20.1]{nS}$,
implying that more than $98\%$ of the voltage falls on barrier $1$.
We shall later see the implications of this asymmetry on the emission
spectrum in these devices, as compared to symmetric devices.

Light emission in the visible range occurs for $V_{SD}>\unit[1.5]{V}$,
much larger than the charging energy of our devices ($\sim\unit[8]{meV}$).
In this regime multiple transmission channels are possible and the
device functions as a simple tunnel junction. The current-voltage
relation at this regime is shown in Figure~\ref{fig2}(b), where we
use the Fowler-Nordheim representation. One can see that at $V_{SD}^{-1}<\unit[0.8]{V^{-1}}$
the curve follows the Fowler-Nordheim equation, $I\sim V_{SD}^{2}\exp\left(-\alpha/V_{SD}\right)$,
which describes tunneling through a trapezoidal barrier ($\alpha$
reflects the barrier parameters). While the high voltage limit is
far beyond the Coulomb blockade regime, the characteristic conductances
values, $g_{1}$ and $g_{2}$, measured at the low voltage still characterize
the voltage division between the two barriers.

\begin{figure}
\includegraphics[width=1\columnwidth]{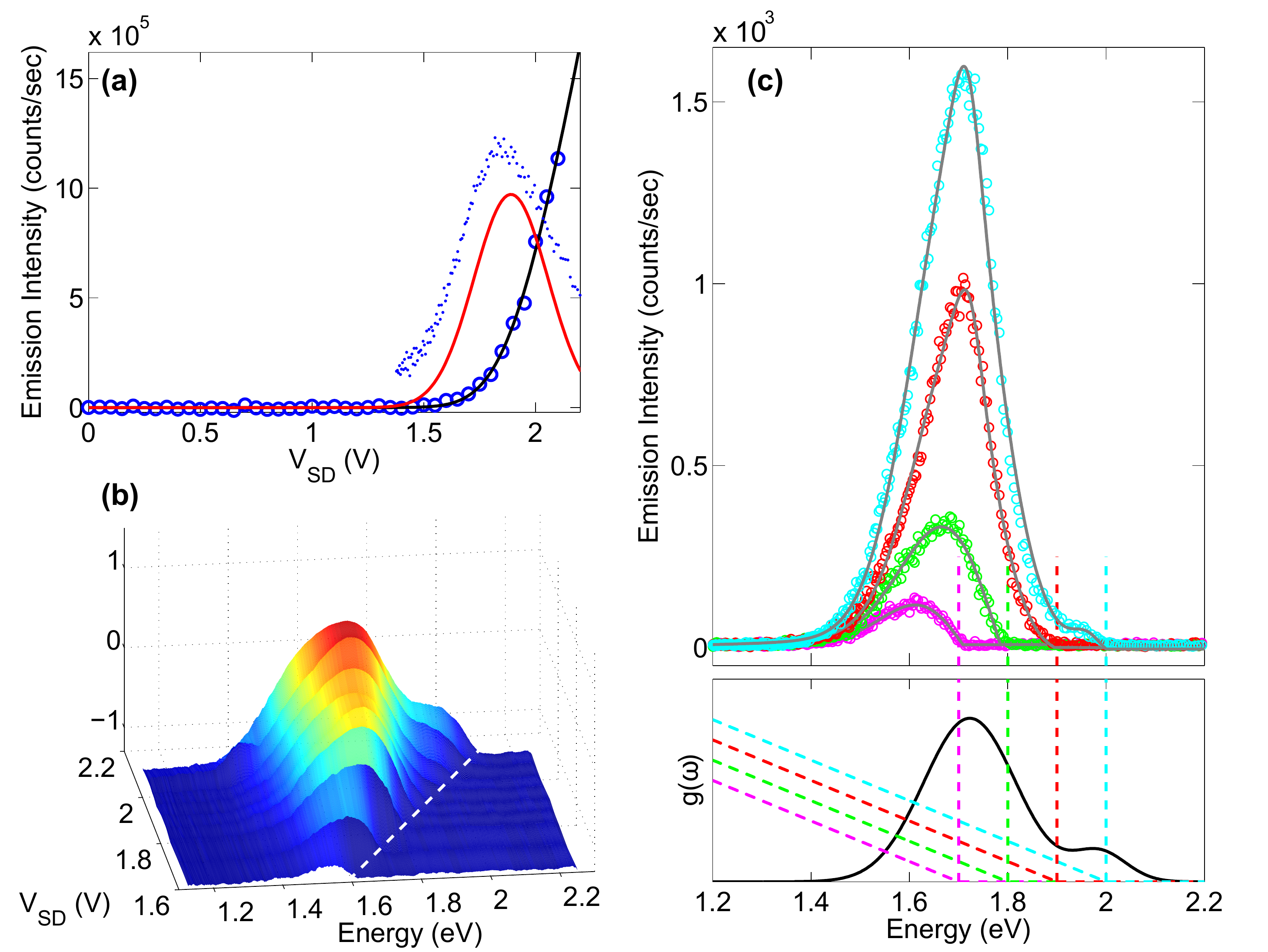}
\caption{(a) Total emitted light intensity, $P\left(V\right)$ (blue circles),
as a function of the applied voltage. The plasmonic spectrum, $g\left(V\right)$,
obtained from the \nth{2} derivative of $P\left(V\right)$ (red line),
and from an independent measurement of the light scattering (blue
dots, shifted) are shown. Error bars are represented by the circles size.
Measurement was done at $T=\unit[6]{K}$.
(b) Spectral evolution of the emitted light
of another device as a function of voltage (logarithmic scale), at
$\unit[50-100]{mV}$ intervals. The gradual build-up of the high energy
part of the plasmonic spectrum is clearly seen, until the cutoff energy
at $\hbar\omega=eV_{SD}$ (white dashed line). (c) Top: emitted light
spectra at $V_{SD}=\unit[1.7,1.8,1.9,2]{V}$ (magenta, green, red
and cyan circles, respectively), fitted to a common plasmonic spectrum
$g\left(\omega\right)$ and different linear excitation power spectra
$\left|I_{\omega}\right|^{2}$ (solid gray lines). Bottom: the common
plasmonic spectrum $g\left(\omega\right)$ used to fit all emitted
spectra (black solid line), and the linear excitation power spectra
with different energy cutoffs at $\hbar\omega=eV_{SD}$ (dashed lines,
colors as in top). The Measurements of b and c were done at $T=\unit[4.5]{K}$.}
\label{fig3}
\end{figure}

Optical measurements were performed using a Nikon Eclipse Ti-E optical
microscope with a $50$X objective (NA=$0.6$), equipped with an Andor
Shamrock $303$i Imaging Spectrograph and a Andor iXon $897$ EMCCD
camera. We find that as $V_{SD}$ increases above $\sim\unit[1.5]{V}$
a bright spot appears at the center of the bow-tie structure, and
becomes brighter as the voltage is further increased. This behavior
is summarized in Figure~\ref{fig3}(a), where the total number of counts
$P$ is plotted as a function of the applied voltage $V_{SD}$ .

The observed rise of $P$ at $V_{SD}>\unit[1.5]{V}$ is due to two
factors: (i) Enhanced tunneling rate, which is linear in $V$, as
given by the current power spectrum $\left|I_{\omega}\right|^{2}$
in eq.~(2). (ii) Plasmonic enhancement
- moving the cutoff voltage across the plasmonic spectrum, $g\left(\omega\right)$.

It is interesting to note that this measurement of $P(V)=\intop_{0}^{eV/\hbar}S(\omega)d\omega$
can directly provide the plasmonic spectrum. It can be easily shown\cite{Lambe1976}
that $\partial^{2}P/\partial V^{2}=g(V)$, hence, by simply measuring
the total emitted light intensity as a function of voltage one can
obtain the plasmonic spectrum of such a structure. Figure~\ref{fig3}(a)
demonstrates the strength of this technique by comparing the spectrum
obtained through this procedure to an independent measurement of the
light scattering spectrum (in order to minimize the numerical noise
we first performed a fit to the curve and then took its numerical
derivative). By performing this procedure on different structures,
with different plasmonic resonances, we have verified that this procedure
indeed yields the correct spectrum\bibnotemark[2]. Clearly, to obtain
fine features in the spectrum one needs to conduct measurements with
high signal-to-noise ratio.

Let us now turn to study the evolution of the emitted spectrum with
voltage. As the voltage increases, the emitted spectrum evolves due
to an increased tunneling rate and a larger overlap of the excitation
energies with the plasmonic spectrum. Figure \ref{fig3}(b) shows
a compilation of $10$ spectra, at $\unit[50-100]{mV}$ intervals
between $\unit[1.6]{V}$ and $\unit[2.15]{V}$, measured by dispersing
the emitted light into a spectrometer (the intensity is presented
on a logarithmic scale). Indeed, one can clearly observe how the high
energy part of the plasmonic spectrum builds up with increasing voltage
and the overall intensity becomes higher. The cutoff energy exactly
at $\hslash\omega=eV_{SD}$ can be clearly seen in the figure. We
find that we could use eq.~(2) to fit all
the spectra, using the same plasmonic spectrum, $g(\omega)$, and
linear excitation power spectrum, $\left|I_{\omega}\right|^{2}$.
This is demonstrated for four spectra in Figure~\ref{fig3}(c).

An implicit assumption that we used in the fits of Figure~\ref{fig3}(c)
was that $V=V_{SD}$, namely - all the voltage that is applied on
the device falls on one barrier. As was shown in Figure~\ref{fig2}(a)
this is indeed the case for very asymmetric devices, where $g_{2}/g_{1}\gg10$,
but is not generally true. Measurements of different, more symmetric
devices, where $g_{2}/g_{1}\thickapprox1$, show a cutoff energy which
is significantly lower than $V_{SD}$, up to a factor of two\bibnotemark[2].
We find that such symmetric devices experience large temporal fluctuations
in the spectral lineshape, and in particular in the cutoff energy.
In these devices small changes in the tunnel junction conductances
may result in a significant different voltage division, and consequently
- different cutoff energy. This emphasizes the importance of the low
voltage characterization at low temperatures.

\begin{figure}
\includegraphics[width=1\columnwidth]{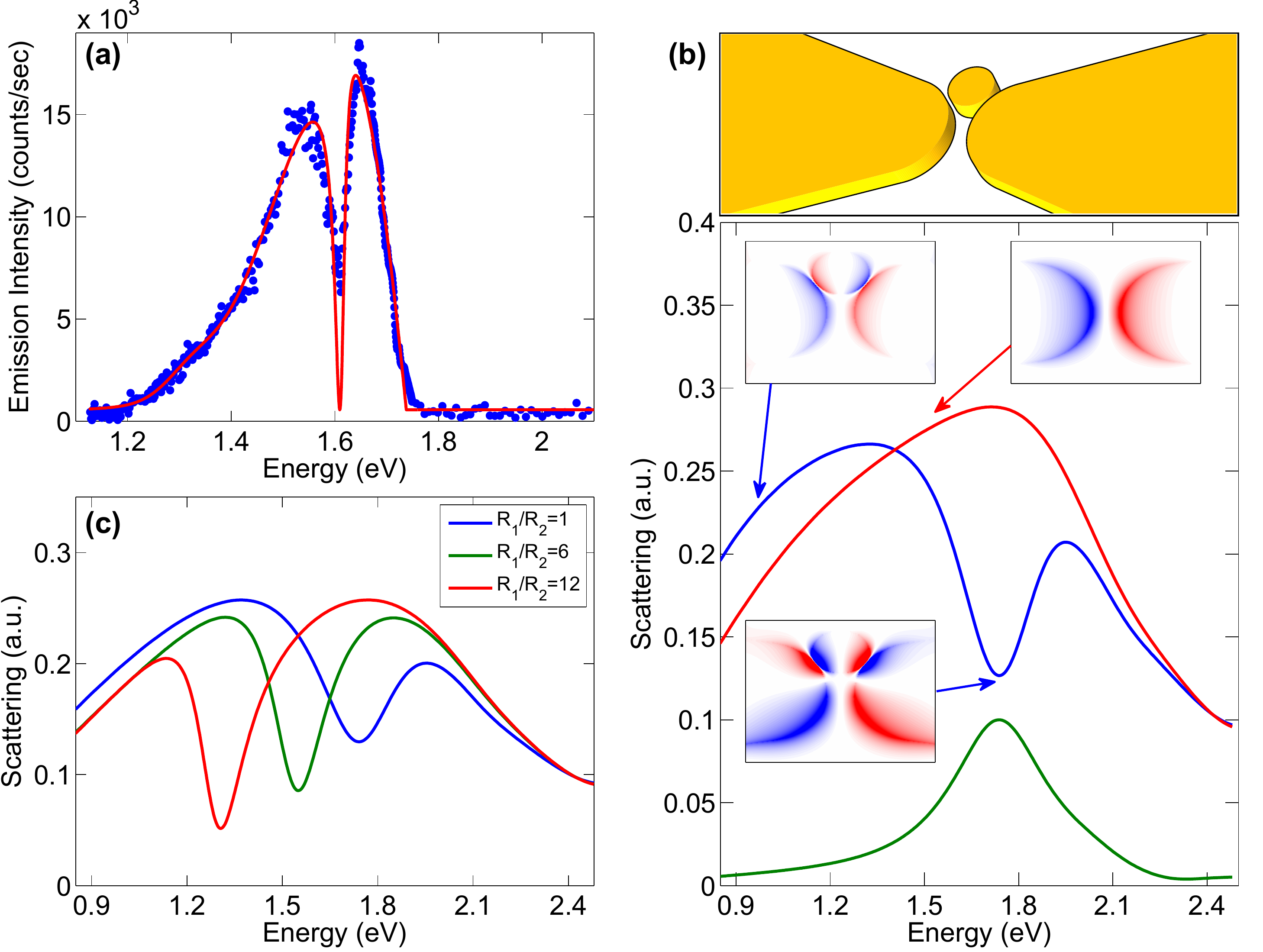}
\caption{(a) Measurement of light emission spectrum in a device demonstrating
a Fano spectral lineshape (blue circles), and a fit with a Fano factor
$F=0.22$ and cutoff energy $\unit[1.74]{eV}$ (red solid line). Measurement was done at $T=\unit[4.5]{K}$.
(b) FDTD simulation of the bow-tie electrodes and a trapped nano-disk.
Top: sketch of the simulated geometry. Bottom: scattering spectra
of electrodes with (blue) and without (red) nano-disk, and of an emitting dipole localized at the nano-disk center (green).
The insets show the spatial charge distribution at the different regimes, represented
by the normal electric field component. (c) FDTD simulation of bow-tie
electrodes and a trapped nano-disk with different eccentricity ($R_{1}$,$R_{2}$
are the major and minor axes of the ellipse), presenting the ability
to control the Fano resonance in our structure.}
\label{fig4}
\end{figure}

An important feature of electrically driven plasmon is the local nature
of the excitation. Contrary to a scattering measurement, in which
plasmons are excited over a macroscopic area, here the plasmons are
generated at the tunnel barrier over which the voltage drops. In that
sense, this method is similar to electron energy loss spectroscopy
(EELS), where a focused beam of electron excites a plasmon locally\cite{Coenen2015}.
The implication is that we are able to probe local features of the
plasmonic spectrum $g(\omega)$. This is well manifested in our bow-tie
structure: while the far field scattering spectrum is dominated by
contributions of the large size electrodes, the local excitation emphasizes
fine features arising from plasmonic interactions in the gap region.
Indeed, in several devices ($3$ out of the $10$ measured) we find a peculiar spectrum, characterized
by two adjacent peaks, with an asymmetric dip between them, as shown
in Figure~\ref{fig4}(a). We obtain a good fit of this spectrum to
a Fano behavior $\left(F\gamma+\omega-\omega_{0}\right)^{2}/\left[\left(\omega-\omega_{0}\right)^{2}+\gamma^{2}\right]$,
with a Fano factor $F=0.22$, indicating a strong interference between
two resonances.

To explain the appearance of a Fano resonance\cite{Fano1961,Miroshnichenko2010}
in our structure we use a simple toy-model, consisting of infinite
electrodes and a metal NP disk between them (top of Figure~\ref{fig4}(b)).
We calculate the plasmonic spectrum of this structure using Finite-difference
time-domain (FDTD) method\bibnotemark[2]. Figure \ref{fig4}(b)
shows the calculated scattering spectrum obtained by exciting a small
region around the disk. This excitation scheme is a simple way to
represent the local nature of the excitation in our experiments. Remarkably, one
obtains in this simple toy-model the clear Fano hallmarks: strong
asymmetry and a pronounced interference dip between the peaks. As
seen, these features appear only when the disk is present. We have
verified that similar spectra are obtained when the disk is replaced
by a sphere.
Our calculations clearly demonstrate the importance of the local nature of the excitation
in the observation of the Fano resonance.
We find that when we expand the illumination area the Fano feature gradually
disappears, and a simple single broad peak remains.
It is seen that the experimental spectral features (scattering and emission, Figures~\ref{fig3},\ref{fig4}(a)) are narrower than those obtained in the simulation.
This difference is explained in detail in the Supporting Information.

To understand how the Fano spectrum is created here it is instructive
to examine the spatial charge distribution at the surface (represented
by the normal electric field component), as shown in Figure~\ref{fig4}(b).
It is seen that without the NP we obtain a simple dipolar distribution
around the gap. This is manifested as a broad spectral lineshape,
which serves as the continuum in the Fano representation\cite{Giannini2011}.
The introduction of the NP radically changes this charge distribution:
the induced dipolar field at the NP forces an overall quadrupole-like
distribution in the structure, which suppresses the far field emission\cite{Hao2008,Wu2011a}.

In order to demonstrate that the dip is indeed due to destructive interference between the NP and the electrodes resonances, we "turn off" this interference by exciting the structure with a dipole, localized at the NP center. This way of excitation affects mainly the NP, and the electrodes are only weakly excited.
The calculated spectrum under this excitation condition shows a clear peak, which is located at exactly the same spectral position as the Fano dip (Figure~\ref{fig4}(b)).
This proves that indeed the electrodes play the role of a continuum, and the NP - of the narrow resonance.

The particle in a bow-tie structure allows easy control of the Fano resonance. This is
demonstrated in Figure~\ref{fig4}(c), where the eccentricity of
the NP is gradually changed, from a circular disk to a narrow ellipse,
while keeping the other structural parameters fixed. Remarkably, the
Fano dip moves to lower energy and becomes narrower, following the corresponding changes
in the NP resonance spectral position and width. On the other hand, the Fano resonance
is relatively insensitive to the details of the electrodes design:
when changing the angle of the bow-tie electrodes we find that the
Fano dip remains at approximately the same spectral position\bibnotemark[2].
It is important to note, however, that as the NP is moved down towards the bow-tie center,
the Fano resonance disappears\cite{Hao2008}. This could explain the fact that the resonance is observed only in some of the devices: we believe
that in these devices the NP was trapped off the bow-tie center.

Fano resonances in plasmonic structures are of great recent interest\cite{Hao2008,Neubrech2008,Aizpurua2008,Verellen2009,Fan2010,Lukyanchuk2010,Miroshnichenko2010,Pryce2010,Giannini2011,Wu2011a}.
It is suggested that their sharp characteristic spectral features
and sensitivity to the sample parameters may turn them into efficient
sensors. In that sense, the ability to generate such resonances in
electrically driven devices may amplify their potential. Furthermore,
the separate control of the interfering broad and narrow modes in
these devices offers an easy way of engineering the Fano resonance.
Such devices may be a step toward the realization of an on-chip, controllable
nano-optical emitters and sensors, and be of a great interest to the
near-field microscopy community.

\begin{suppinfo}
Nanoparticle synthesis and device fabrication methods; Single dot
simulation; Symmetrical and asymmetrical conductance devices; FDTD
simulation details; Fano resonance measurements;
\end{suppinfo}

The authors declare no competing financial interest.

\begin{acknowledgement}
We would like to thank D. Mahalu and O. Raslin for their help in the
electron beam lithography.
\end{acknowledgement}

\bibnotetext[2]{See Supporting Information}

\bibliography{ElectricalDrivenPlasmon}

\providecommand{\latin}[1]{#1}
\providecommand*\mcitethebibliography{\thebibliography}
\csname @ifundefined\endcsname{endmcitethebibliography}
  {\let\endmcitethebibliography\endthebibliography}{}
\begin{mcitethebibliography}{41}
\providecommand*\natexlab[1]{#1}
\providecommand*\mciteSetBstSublistMode[1]{}
\providecommand*\mciteSetBstMaxWidthForm[2]{}
\providecommand*\mciteBstWouldAddEndPuncttrue
  {\def\EndOfBibitem{\unskip.}}
\providecommand*\mciteBstWouldAddEndPunctfalse
  {\let\EndOfBibitem\relax}
\providecommand*\mciteSetBstMidEndSepPunct[3]{}
\providecommand*\mciteSetBstSublistLabelBeginEnd[3]{}
\providecommand*\EndOfBibitem{}
\mciteSetBstSublistMode{f}
\mciteSetBstMaxWidthForm{subitem}{(\alph{mcitesubitemcount})}
\mciteSetBstSublistLabelBeginEnd
  {\mcitemaxwidthsubitemform\space}
  {\relax}
  {\relax}

\bibitem[{Garc{\'{\i}}a de Abajo}(2010)]{GarciadeAbajo2010}
{Garc{\'{\i}}a de Abajo},~F.~J. \emph{Rev. Mod. Phys.} \textbf{2010},
  \emph{82}, 209--275\relax
\mciteBstWouldAddEndPuncttrue
\mciteSetBstMidEndSepPunct{\mcitedefaultmidpunct}
{\mcitedefaultendpunct}{\mcitedefaultseppunct}\relax
\EndOfBibitem
\bibitem[Fan \latin{et~al.}(2012)Fan, Colombo, Huang, Krogstrup, Nyg{\aa}rd,
  {Fontcuberta i Morral}, and Brongersma]{Fan2012}
Fan,~P.; Colombo,~C.; Huang,~K. C.~Y.; Krogstrup,~P.; Nyg{\aa}rd,~J.;
  {Fontcuberta i Morral},~A.; Brongersma,~M.~L. \emph{Nano Lett.}
  \textbf{2012}, \emph{12}, 4943--4947\relax
\mciteBstWouldAddEndPuncttrue
\mciteSetBstMidEndSepPunct{\mcitedefaultmidpunct}
{\mcitedefaultendpunct}{\mcitedefaultseppunct}\relax
\EndOfBibitem
\bibitem[Li and Stockman(2013)Li, and Stockman]{Li2013}
Li,~D.; Stockman,~M.~I. \emph{Phys. Rev. Lett.} \textbf{2013}, \emph{110},
  106803\relax
\mciteBstWouldAddEndPuncttrue
\mciteSetBstMidEndSepPunct{\mcitedefaultmidpunct}
{\mcitedefaultendpunct}{\mcitedefaultseppunct}\relax
\EndOfBibitem
\bibitem[Rai \latin{et~al.}(2013)Rai, Hartmann, Berthelot, Arocas, {Colas des
  Francs}, Hartschuh, and Bouhelier]{Rai2013}
Rai,~P.; Hartmann,~N.; Berthelot,~J.; Arocas,~J.; {Colas des Francs},~G.;
  Hartschuh,~A.; Bouhelier,~A. \emph{Phys. Rev. Lett.} \textbf{2013},
  \emph{111}, 026804\relax
\mciteBstWouldAddEndPuncttrue
\mciteSetBstMidEndSepPunct{\mcitedefaultmidpunct}
{\mcitedefaultendpunct}{\mcitedefaultseppunct}\relax
\EndOfBibitem
\bibitem[Sheldon \latin{et~al.}(2014)Sheldon, van~de Groep, Brown, Polman, and
  Atwater]{Sheldon2014}
Sheldon,~M.~T.; van~de Groep,~J.; Brown,~A.~M.; Polman,~A.; Atwater,~H.~A.
  \emph{Science} \textbf{2014}, \emph{346}, 828--831\relax
\mciteBstWouldAddEndPuncttrue
\mciteSetBstMidEndSepPunct{\mcitedefaultmidpunct}
{\mcitedefaultendpunct}{\mcitedefaultseppunct}\relax
\EndOfBibitem
\bibitem[Savage \latin{et~al.}(2012)Savage, Hawkeye, Esteban, Borisov,
  Aizpurua, and Baumberg]{Savage2012}
Savage,~K.~J.; Hawkeye,~M.~M.; Esteban,~R.; Borisov,~A.~G.; Aizpurua,~J.;
  Baumberg,~J.~J. \emph{Nature} \textbf{2012}, \emph{491}, 574--577\relax
\mciteBstWouldAddEndPuncttrue
\mciteSetBstMidEndSepPunct{\mcitedefaultmidpunct}
{\mcitedefaultendpunct}{\mcitedefaultseppunct}\relax
\EndOfBibitem
\bibitem[Hao \latin{et~al.}(2008)Hao, Sonnefraud, {Van Dorpe}, Maier, Halas,
  and Nordlander]{Hao2008}
Hao,~F.; Sonnefraud,~Y.; {Van Dorpe},~P.; Maier,~S.~a.; Halas,~N.~J.;
  Nordlander,~P. \emph{Nano Lett.} \textbf{2008}, \emph{8}, 3983--3988\relax
\mciteBstWouldAddEndPuncttrue
\mciteSetBstMidEndSepPunct{\mcitedefaultmidpunct}
{\mcitedefaultendpunct}{\mcitedefaultseppunct}\relax
\EndOfBibitem
\bibitem[Benda{\~{n}}a \latin{et~al.}(2011)Benda{\~{n}}a, Polman, and
  {Garc{\'{\i}}a de Abajo}]{Bendana2011}
Benda{\~{n}}a,~X.; Polman,~A.; {Garc{\'{\i}}a de Abajo},~F.~J. \emph{Nano
  Lett.} \textbf{2011}, \emph{11}, 5099--5103\relax
\mciteBstWouldAddEndPuncttrue
\mciteSetBstMidEndSepPunct{\mcitedefaultmidpunct}
{\mcitedefaultendpunct}{\mcitedefaultseppunct}\relax
\EndOfBibitem
\bibitem[Jeong \latin{et~al.}(2013)Jeong, No, Hwang, Kim, Seo, Park, and
  Lee]{Jeong2013}
Jeong,~K.-Y.; No,~Y.-S.; Hwang,~Y.; Kim,~K.~S.; Seo,~M.-K.; Park,~H.-G.;
  Lee,~Y.-H. \emph{Nat. Commun.} \textbf{2013}, \emph{4}, 1--6\relax
\mciteBstWouldAddEndPuncttrue
\mciteSetBstMidEndSepPunct{\mcitedefaultmidpunct}
{\mcitedefaultendpunct}{\mcitedefaultseppunct}\relax
\EndOfBibitem
\bibitem[Huang \latin{et~al.}(2014)Huang, Seo, Sarmiento, Huo, Harris, and
  Brongersma]{Huang2014}
Huang,~K. C.~Y.; Seo,~M.-K.; Sarmiento,~T.; Huo,~Y.; Harris,~J.~S.;
  Brongersma,~M.~L. \emph{Nat. Photonics} \textbf{2014}, \emph{8},
  244--249\relax
\mciteBstWouldAddEndPuncttrue
\mciteSetBstMidEndSepPunct{\mcitedefaultmidpunct}
{\mcitedefaultendpunct}{\mcitedefaultseppunct}\relax
\EndOfBibitem
\bibitem[Cox and {Javier Garc{\'{\i}}a de Abajo}(2014)Cox, and {Javier
  Garc{\'{\i}}a de Abajo}]{Cox2014}
Cox,~J.~D.; {Javier Garc{\'{\i}}a de Abajo},~F. \emph{Nat. Commun.}
  \textbf{2014}, \emph{5}, 5725\relax
\mciteBstWouldAddEndPuncttrue
\mciteSetBstMidEndSepPunct{\mcitedefaultmidpunct}
{\mcitedefaultendpunct}{\mcitedefaultseppunct}\relax
\EndOfBibitem
\bibitem[Tielrooij \latin{et~al.}(2015)Tielrooij, Orona, Ferrier, Badioli,
  Navickaite, Coop, Nanot, Kalinic, Cesca, Gaudreau, Ma, Centeno, Pesquera,
  Zurutuza, de~Riedmatten, Goldner, {Garc{\'{\i}}a de Abajo}, Jarillo-Herrero,
  and Koppens]{Tielrooij2015}
Tielrooij,~K.~J.; Orona,~L.; Ferrier,~A.; Badioli,~M.; Navickaite,~G.;
  Coop,~S.; Nanot,~S.; Kalinic,~B.; Cesca,~T.; Gaudreau,~L. \latin{et~al.}
  \emph{Nat. Phys.} \textbf{2015}, \emph{11}, 281--287\relax
\mciteBstWouldAddEndPuncttrue
\mciteSetBstMidEndSepPunct{\mcitedefaultmidpunct}
{\mcitedefaultendpunct}{\mcitedefaultseppunct}\relax
\EndOfBibitem
\bibitem[Lambe and McCarthy(1976)Lambe, and McCarthy]{Lambe1976}
Lambe,~J.; McCarthy,~S.~L. \emph{Phys. Rev. Lett.} \textbf{1976}, \emph{37},
  923--925\relax
\mciteBstWouldAddEndPuncttrue
\mciteSetBstMidEndSepPunct{\mcitedefaultmidpunct}
{\mcitedefaultendpunct}{\mcitedefaultseppunct}\relax
\EndOfBibitem
\bibitem[Hansma and Broida(1978)Hansma, and Broida]{Hansma1978}
Hansma,~P.~K.; Broida,~H.~P. \emph{Appl. Phys. Lett.} \textbf{1978}, \emph{32},
  545\relax
\mciteBstWouldAddEndPuncttrue
\mciteSetBstMidEndSepPunct{\mcitedefaultmidpunct}
{\mcitedefaultendpunct}{\mcitedefaultseppunct}\relax
\EndOfBibitem
\bibitem[Hone \latin{et~al.}(1978)Hone, Muhlschlegel, and Scalapino]{Hone1978}
Hone,~D.; Muhlschlegel,~B.; Scalapino,~D.~J. \emph{Appl. Phys. Lett.}
  \textbf{1978}, \emph{33}, 203\relax
\mciteBstWouldAddEndPuncttrue
\mciteSetBstMidEndSepPunct{\mcitedefaultmidpunct}
{\mcitedefaultendpunct}{\mcitedefaultseppunct}\relax
\EndOfBibitem
\bibitem[Rendell and Scalapino(1981)Rendell, and Scalapino]{Rendell1981}
Rendell,~R.~W.; Scalapino,~D.~J. \emph{Phys. Rev. B} \textbf{1981}, \emph{24},
  3276--3294\relax
\mciteBstWouldAddEndPuncttrue
\mciteSetBstMidEndSepPunct{\mcitedefaultmidpunct}
{\mcitedefaultendpunct}{\mcitedefaultseppunct}\relax
\EndOfBibitem
\bibitem[Coombs \latin{et~al.}(1988)Coombs, Gimzewski, Reihl, Sass, and
  Schlittler]{Coombs1988}
Coombs,~J.~H.; Gimzewski,~J.~K.; Reihl,~B.; Sass,~J.~K.; Schlittler,~R.~R.
  \emph{J. Microsc. (Oxford, U. K.)} \textbf{1988}, \emph{152}, 325--336\relax
\mciteBstWouldAddEndPuncttrue
\mciteSetBstMidEndSepPunct{\mcitedefaultmidpunct}
{\mcitedefaultendpunct}{\mcitedefaultseppunct}\relax
\EndOfBibitem
\bibitem[Gimzewski \latin{et~al.}(1989)Gimzewski, Sass, Schlitter, and
  Schott]{Gimzewski1989}
Gimzewski,~J.~K.; Sass,~J.~K.; Schlitter,~R.~R.; Schott,~J. \emph{Europhys.
  Lett.} \textbf{1989}, \emph{8}, 435--440\relax
\mciteBstWouldAddEndPuncttrue
\mciteSetBstMidEndSepPunct{\mcitedefaultmidpunct}
{\mcitedefaultendpunct}{\mcitedefaultseppunct}\relax
\EndOfBibitem
\bibitem[Berndt \latin{et~al.}(1991)Berndt, Gimzewski, and
  Johansson]{Berndt1991}
Berndt,~R.; Gimzewski,~J.~K.; Johansson,~P. \emph{Phys. Rev. Lett.}
  \textbf{1991}, \emph{67}, 3796--3799\relax
\mciteBstWouldAddEndPuncttrue
\mciteSetBstMidEndSepPunct{\mcitedefaultmidpunct}
{\mcitedefaultendpunct}{\mcitedefaultseppunct}\relax
\EndOfBibitem
\bibitem[Kern \latin{et~al.}(2015)Kern, Kullock, Prangsma, Emmerling, Kamp, and
  Hecht]{Kern2015}
Kern,~J.; Kullock,~R.; Prangsma,~J.; Emmerling,~M.; Kamp,~M.; Hecht,~B.
  \emph{Nat. Photonics} \textbf{2015}, \emph{9}, 582--586\relax
\mciteBstWouldAddEndPuncttrue
\mciteSetBstMidEndSepPunct{\mcitedefaultmidpunct}
{\mcitedefaultendpunct}{\mcitedefaultseppunct}\relax
\EndOfBibitem
\bibitem[Persson and Baratoff(1992)Persson, and Baratoff]{Persson1992}
Persson,~B. N.~J.; Baratoff,~A. \emph{Phys. Rev. Lett.} \textbf{1992},
  \emph{68}, 3224--3227\relax
\mciteBstWouldAddEndPuncttrue
\mciteSetBstMidEndSepPunct{\mcitedefaultmidpunct}
{\mcitedefaultendpunct}{\mcitedefaultseppunct}\relax
\EndOfBibitem
\bibitem[Uehara \latin{et~al.}(1992)Uehara, Kimura, Ushioda, and
  Takeuchi]{Uehara1992}
Uehara,~Y.; Kimura,~Y.; Ushioda,~S.; Takeuchi,~K. \emph{Jpn. J. Appl. Phys.}
  \textbf{1992}, \emph{31}, 2465--2469\relax
\mciteBstWouldAddEndPuncttrue
\mciteSetBstMidEndSepPunct{\mcitedefaultmidpunct}
{\mcitedefaultendpunct}{\mcitedefaultseppunct}\relax
\EndOfBibitem
\bibitem[Schneider \latin{et~al.}(2013)Schneider, Johansson, and
  Berndt]{Schneider2013}
Schneider,~N.~L.; Johansson,~P.; Berndt,~R. \emph{Phys. Rev. B} \textbf{2013},
  \emph{87}, 045409\relax
\mciteBstWouldAddEndPuncttrue
\mciteSetBstMidEndSepPunct{\mcitedefaultmidpunct}
{\mcitedefaultendpunct}{\mcitedefaultseppunct}\relax
\EndOfBibitem
\bibitem[Kaasbjerg and Nitzan(2015)Kaasbjerg, and Nitzan]{Kaasbjerg2015}
Kaasbjerg,~K.; Nitzan,~A. \emph{Phys. Rev. Lett.} \textbf{2015}, \emph{114},
  1--6\relax
\mciteBstWouldAddEndPuncttrue
\mciteSetBstMidEndSepPunct{\mcitedefaultmidpunct}
{\mcitedefaultendpunct}{\mcitedefaultseppunct}\relax
\EndOfBibitem
\bibitem[Buret \latin{et~al.}(2015)Buret, Uskov, Dellinger, Cazier,
  Mennemanteuil, Berthelot, Smetanin, Protsenko, Colas-des Francs, and
  Bouhelier]{Buret2015}
Buret,~M.; Uskov,~A.~V.; Dellinger,~J.; Cazier,~N.; Mennemanteuil,~M.-M.;
  Berthelot,~J.; Smetanin,~I.~V.; Protsenko,~I.~E.; Colas-des Francs,~G.;
  Bouhelier,~A. \emph{Nano Lett.} \textbf{2015}, \emph{15}, 5811--5818\relax
\mciteBstWouldAddEndPuncttrue
\mciteSetBstMidEndSepPunct{\mcitedefaultmidpunct}
{\mcitedefaultendpunct}{\mcitedefaultseppunct}\relax
\EndOfBibitem
\bibitem[Guttman \latin{et~al.}(2011)Guttman, Mahalu, Sperling, Cohen-Hoshen,
  and Bar-Joseph]{Guttman2011}
Guttman,~A.; Mahalu,~D.; Sperling,~J.; Cohen-Hoshen,~E.; Bar-Joseph,~I.
  \emph{Appl. Phys. Lett.} \textbf{2011}, \emph{99}, 063113\relax
\mciteBstWouldAddEndPuncttrue
\mciteSetBstMidEndSepPunct{\mcitedefaultmidpunct}
{\mcitedefaultendpunct}{\mcitedefaultseppunct}\relax
\EndOfBibitem
\bibitem[Vardi \latin{et~al.}(2014)Vardi, Guttman, and Bar-Joseph]{Vardi2014}
Vardi,~Y.; Guttman,~A.; Bar-Joseph,~I. \emph{Nano Lett.} \textbf{2014},
  \emph{14}, 2794--2799\relax
\mciteBstWouldAddEndPuncttrue
\mciteSetBstMidEndSepPunct{\mcitedefaultmidpunct}
{\mcitedefaultendpunct}{\mcitedefaultseppunct}\relax
\EndOfBibitem
\bibitem[Cohen-Hoshen \latin{et~al.}(2012)Cohen-Hoshen, Bryant, Pinkas,
  Sperling, and Bar-Joseph]{Cohen-Hoshen2012}
Cohen-Hoshen,~E.; Bryant,~G.~W.; Pinkas,~I.; Sperling,~J.; Bar-Joseph,~I.
  \emph{Nano Lett.} \textbf{2012}, \emph{12}, 4260--4\relax
\mciteBstWouldAddEndPuncttrue
\mciteSetBstMidEndSepPunct{\mcitedefaultmidpunct}
{\mcitedefaultendpunct}{\mcitedefaultseppunct}\relax
\EndOfBibitem
\bibitem[2()]{2}
See Supporting Information\relax
\mciteBstWouldAddEndPuncttrue
\mciteSetBstMidEndSepPunct{\mcitedefaultmidpunct}
{\mcitedefaultendpunct}{\mcitedefaultseppunct}\relax
\EndOfBibitem
\bibitem[Coenen \latin{et~al.}(2015)Coenen, Schoen, Mann, Rodriguez, Brenny,
  Polman, and Brongersma]{Coenen2015}
Coenen,~T.; Schoen,~D.~T.; Mann,~S.~A.; Rodriguez,~S. R.~K.; Brenny,~B. J.~M.;
  Polman,~A.; Brongersma,~M.~L. \emph{Nano Lett.} \textbf{2015},
  151016141704002\relax
\mciteBstWouldAddEndPuncttrue
\mciteSetBstMidEndSepPunct{\mcitedefaultmidpunct}
{\mcitedefaultendpunct}{\mcitedefaultseppunct}\relax
\EndOfBibitem
\bibitem[Fano(1961)]{Fano1961}
Fano,~U. \emph{Phys. Rev.} \textbf{1961}, \emph{124}, 1866--1878\relax
\mciteBstWouldAddEndPuncttrue
\mciteSetBstMidEndSepPunct{\mcitedefaultmidpunct}
{\mcitedefaultendpunct}{\mcitedefaultseppunct}\relax
\EndOfBibitem
\bibitem[Miroshnichenko \latin{et~al.}(2010)Miroshnichenko, Flach, and
  Kivshar]{Miroshnichenko2010}
Miroshnichenko,~A.~E.; Flach,~S.; Kivshar,~Y.~S. \emph{Rev. Mod. Phys.}
  \textbf{2010}, \emph{82}, 2257--2298\relax
\mciteBstWouldAddEndPuncttrue
\mciteSetBstMidEndSepPunct{\mcitedefaultmidpunct}
{\mcitedefaultendpunct}{\mcitedefaultseppunct}\relax
\EndOfBibitem
\bibitem[Giannini \latin{et~al.}(2011)Giannini, Francescato, Amrania, Phillips,
  and Maier]{Giannini2011}
Giannini,~V.; Francescato,~Y.; Amrania,~H.; Phillips,~C.~C.; Maier,~S.~A.
  \emph{Nano Lett.} \textbf{2011}, \emph{11}, 2835--2840\relax
\mciteBstWouldAddEndPuncttrue
\mciteSetBstMidEndSepPunct{\mcitedefaultmidpunct}
{\mcitedefaultendpunct}{\mcitedefaultseppunct}\relax
\EndOfBibitem
\bibitem[Wu \latin{et~al.}(2011)Wu, Khanikaev, Adato, Arju, Yanik, Altug, and
  Shvets]{Wu2011a}
Wu,~C.; Khanikaev,~A.~B.; Adato,~R.; Arju,~N.; Yanik,~A.~A.; Altug,~H.;
  Shvets,~G. \emph{Nat. Mater.} \textbf{2011}, \emph{11}, 69--75\relax
\mciteBstWouldAddEndPuncttrue
\mciteSetBstMidEndSepPunct{\mcitedefaultmidpunct}
{\mcitedefaultendpunct}{\mcitedefaultseppunct}\relax
\EndOfBibitem
\bibitem[Neubrech \latin{et~al.}(2008)Neubrech, Pucci, Cornelius, Karim,
  Garc{\'{\i}}a-Etxarri, and Aizpurua]{Neubrech2008}
Neubrech,~F.; Pucci,~A.; Cornelius,~T.~W.; Karim,~S.;
  Garc{\'{\i}}a-Etxarri,~A.; Aizpurua,~J. \emph{Phys. Rev. Lett.}
  \textbf{2008}, \emph{101}, 157403\relax
\mciteBstWouldAddEndPuncttrue
\mciteSetBstMidEndSepPunct{\mcitedefaultmidpunct}
{\mcitedefaultendpunct}{\mcitedefaultseppunct}\relax
\EndOfBibitem
\bibitem[Aizpurua \latin{et~al.}(2008)Aizpurua, Taubner, {Garc{\'{\i}}a de
  Abajo}, Brehm, and Hillenbrand]{Aizpurua2008}
Aizpurua,~J.; Taubner,~T.; {Garc{\'{\i}}a de Abajo},~F.~J.; Brehm,~M.;
  Hillenbrand,~R. \emph{Opt. Express} \textbf{2008}, \emph{16}, 1529\relax
\mciteBstWouldAddEndPuncttrue
\mciteSetBstMidEndSepPunct{\mcitedefaultmidpunct}
{\mcitedefaultendpunct}{\mcitedefaultseppunct}\relax
\EndOfBibitem
\bibitem[Verellen \latin{et~al.}(2009)Verellen, Sonnefraud, Sobhani, Hao,
  Moshchalkov, Dorpe, Nordlander, and Maier]{Verellen2009}
Verellen,~N.; Sonnefraud,~Y.; Sobhani,~H.; Hao,~F.; Moshchalkov,~V.~V.;
  Dorpe,~P.~V.; Nordlander,~P.; Maier,~S.~A. \emph{Nano Lett.} \textbf{2009},
  \emph{9}, 1663--1667\relax
\mciteBstWouldAddEndPuncttrue
\mciteSetBstMidEndSepPunct{\mcitedefaultmidpunct}
{\mcitedefaultendpunct}{\mcitedefaultseppunct}\relax
\EndOfBibitem
\bibitem[Fan \latin{et~al.}(2010)Fan, Wu, Bao, Bao, Bardhan, Halas, Manoharan,
  Nordlander, Shvets, and Capasso]{Fan2010}
Fan,~J.~a.; Wu,~C.; Bao,~K.; Bao,~J.; Bardhan,~R.; Halas,~N.~J.;
  Manoharan,~V.~N.; Nordlander,~P.; Shvets,~G.; Capasso,~F. \emph{Science}
  \textbf{2010}, \emph{328}, 1135--1138\relax
\mciteBstWouldAddEndPuncttrue
\mciteSetBstMidEndSepPunct{\mcitedefaultmidpunct}
{\mcitedefaultendpunct}{\mcitedefaultseppunct}\relax
\EndOfBibitem
\bibitem[Luk'yanchuk \latin{et~al.}(2010)Luk'yanchuk, Zheludev, Maier, Halas,
  Nordlander, Giessen, and Chong]{Lukyanchuk2010}
Luk'yanchuk,~B.; Zheludev,~N.~I.; Maier,~S.~A.; Halas,~N.~J.; Nordlander,~P.;
  Giessen,~H.; Chong,~C.~T. \emph{Nat. Mater.} \textbf{2010}, \emph{9},
  707--715\relax
\mciteBstWouldAddEndPuncttrue
\mciteSetBstMidEndSepPunct{\mcitedefaultmidpunct}
{\mcitedefaultendpunct}{\mcitedefaultseppunct}\relax
\EndOfBibitem
\bibitem[Pryce \latin{et~al.}(2010)Pryce, Aydin, Kelaita, Briggs, and
  Atwater]{Pryce2010}
Pryce,~I.~M.; Aydin,~K.; Kelaita,~Y.~A.; Briggs,~R.~M.; Atwater,~H.~A.
  \emph{Nano Lett.} \textbf{2010}, \emph{10}, 4222--4227\relax
\mciteBstWouldAddEndPuncttrue
\mciteSetBstMidEndSepPunct{\mcitedefaultmidpunct}
{\mcitedefaultendpunct}{\mcitedefaultseppunct}\relax
\EndOfBibitem
\end{mcitethebibliography}

\end{document}